# Distributed network for measuring climatic parameters in heterogeneous environments: application in a greenhouse

Javier López-Martínez[a], José L. Blanco-Claraco[a], José Pérez-Alonso[a], Ángel J. Callejón-Ferre[a,b*]

[a]University of Almería, Department of Engineering, Agrifood Campus of International Excellence (CeiA3), Ctra. de Sacramento s/n, 04120, Almería, Spain.

[b]Laboratory-Observatory Andalusian Working Conditions in the Agricultural Sector (LASA). C/ Albert Einstein s/n. 2ª planta. Isla de la Cartuja, 41092 Sevilla, Spain.

E-mails: jlm167@ual.es, jlblanco@ual.es, jpalonso@ual.es, acallejo@ual.es

* Corresponding author: Tel. +34 950 214 237; Fax: +34 950 015 491

E-mail address: acallejo@ual.es (A.J. Callejón-Ferre)

**Abstract**

In Mediterranean countries of Southern Europe, the climatic conditions are usually favourable to cultivate greenhouse vegetables but not always for workers. The aim of this study was to design a network of weather stations capable of gathering data of environmental parameters related to the wellbeing of workers in greenhouses in south-eastern Spain. The unevenness of the thermal environment was studied both vertically as well as horizontally following guideline ISO 7726. The results indicate that the greenhouse should be considered a heterogeneous environment, implying that, for an evaluation of the environmental conditions related to thermal stress of the workers inside the greenhouse, measurements should be taken at different points within the greenhouse at three heights (ankle, abdomen, and head).

# 1. Introduction

Labourers interact directly with their working environment, which, depending on different production sectors (Figure 1), can vary and should be monitored for dangers and risks to worker health and safety (ILO, 1985).

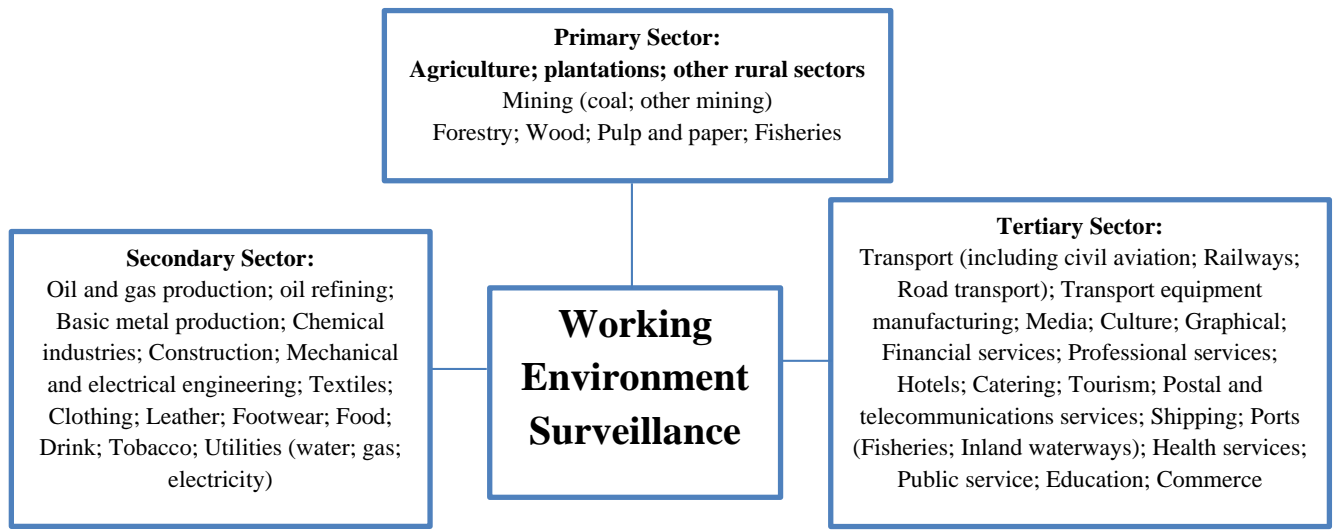

Figure 1. Surveillance of the working environment of different economic sectors.

Monitoring of the working environment is meant to combine all the disciplines related to safety (health, accidents, psychosocial factors, ergonomics, hygiene, illnesses, etc.). In agriculture (primary sector), three factors are distinguished in the discipline of environmental ergonomics: physical factors, organismic factors, and adaptive factors (Rohles, 1985). These factors have greater or lesser implications for the workers according to the tasks undertaken.

Agriculture around the planet is the second source of world employment. It involves contact with plants, animals, machinery, fertilizers, biocides, pests, etc. both in enclosed areas as well as in the open air (ILO, 2010). South-eastern Spain (Almería) has the greatest surface area of greenhouses in Europe, with 30,230 ha directly employing 55,813 workers (per year) of different nationalities (Cabrera-Sánchez et al., 2016).

These greenhouses, lightly built metal structures covered with transparent plastic, have wall and ceiling ventilation with interior diffuse solar radiation. These conditions are favourable for cultivating vegetables (Pérez-Alonso et al., 2011), but are not always suitable for the wellbeing of the greenhouse workers (Callejón-Ferre et al., 2011a). For purely agricultural control, greenhouses are equipped with sensors such as psychrometers, thermometres, pyrometres, conductivimetres, pH metres, etc. (Castilla, 2005).

The vegetables cultivated in the province of Almería are tomatoes, peppers, cucumbers, green beans, eggplant, squash, melons, and watermelons. Workers tend all the tasks over the growing cycle of the vegetables, such as transplanting, pruning, biocide application, harvesting, etc. These tasks last for a complete crop cycle (growing season), i.e. from the end of July to the beginning of June of the following year. Also, maintenance work is carried out when the greenhouse contains no crop (Callejón-Ferre et al., 2009; 2011b).

The greenhouses of south-eastern Spain rarely need heating systems. Nevertheless, the temperature range varies sharply over the four seasons, from 40ºC in the summer to hardly more than 2-3ºC in winter (at night) (Castilla, 2005; Cecchini et al., 2010). Under these conditions, the study of heat stress in humans requires an analysis of the physical magnitudes associated with the environment (temperature, humidity, air velocity, etc.), with the individual, and with the type of work (metabolic rate, acclimation, physical activity, clothes, etc.) (ISO 7933, 2004).

Related to the heat stress in greenhouse climatic conditions, Callejon-Ferre et al. (2011a) studying the thermal conditions of workers in Almería-type greenhouses stated that, during the warmer months, the conditions under which heat stress risk could appear were common. Cecchini et al. (2010) showed that, in greenhouses of central

Italy, the risk for workers during manual harvesting should not be underestimated, pointing to the possibility of the risk of heat stroke during the hottest hours of the day in spring and summer. Also in central Italy, Marucci et al. (2012) studied the heat stress suffered by workers in vegetable grafting greenhouses, concluding that workers were subject to the risk of heat stress mainly between April and October. Similar results have been found for greenhouses of Calabria (southern Italy) (Diano et al., 2016) and Japan (Okushima et al., 2001), where the summertime proved dangerous, especially in the midday. While the risk of heat stress during the hottest months has been pointed, risk of cold stress has also been reported (Callejón-Ferre et al., 2011a). Other studies have examined the heat stress of workers during the greenhouse construction (Pérez-Alonso et al., 2011). This latter scenario differs from the previous ones in that most of the work is done outdoors. The results of above tasks are based on the evaluation of comfort and heat-stress indices.

The thermal-environment assessment is regulated by several rules (Parsons, 2013). The International Standard Organization (ISO) classifies the environment in two categories: moderate and extreme. Each category is evaluated through an appropriate index and the corresponding ISO Standard. Moderate environments should be evaluated through the Predicted Mean Vote index (PMV) according to ISO 7730 Standard (ISO 7730, 2005). Hot extreme environments should be initially treated by means of the Wet Bulb Globe Temperature index (WBGT) as stated at ISO 7243 Standard (ISO 7243, 1985); if limits of the WBGT index are surpassed, a more detailed analysis based on the energy balance equation (subject-environment heat transfer) is required. This analysis must be made according to the ISO 7933 Standard (ISO 7933, 2004), where the Predicted Heat Strain index (PHS) is suggested. Besides ISO standards, other comfort and stress indices have been proposed in the literature for the thermal environment (Epstein and Moran, 2016; D'Ambrosio-Alfano et al., 2011).

The WBGT index, according to ISO 7243 (1985), should be calculated according to one of the next two scenarios: i) inside buildings and outside buildings without a solar load, and ii) outside buildings with a solar load. Since the cover of the greenhouses consists of plastic film (with several additives), workers inside the greenhouse are exposed to diffuse radiation (Nijskens et al., 1985). Thus, conditions inside greenhouses do not exactly match any of the scenarios where WBGT index equation is defined (Callejón-Ferre et al., 2011a). Since the WBGT index might not be adequate for greenhouses, and due to the limitations of this index when the relative humidity is high and the wind speed is low (Budd, 2008), Callejon-Ferre et al. (2011a) used the Humidex Index (HI) (Masterton and Richardson, 1979) instead of WBGT index.

The calculation of the previous indices requires the measurement of several climatic parameters (air temperature, black globe temperature, air velocity, humidity, etc.) and, for PMV and PHS indices, also the metabolic rate related to the worker's physical activity, based in ISO 8996 (2004), and the clothing insulation and sweat rate (ISO 9920, 2007). Measurement of climatic parameters should be conducted in accordance with the ISO 7726 Standard (ISO 7726, 1998). This Standard, establishes the minimum characteristics of instruments for measuring the physical quantities that define a thermal environment. Also, ISO 7726 (1998) refers to the measuring methods, which should take into account that the values of physical quantities may vary in space and time. In case of heterogeneous environments, physical quantities need to be measured at different locations throughout the work place, in a horizontal direction and also in a vertical one. Previous studies address the distribution of air temperature inside greenhouses (López et al., 2012a; 2012b; López et al., 2013; Molina-Aiz et al., 2004;

Granados et al., 2016). Experimental works took place in multispan greenhouses showed maximum differences of air temperature in a horizontal plane, at 1.75 m height from the ground, between 2 ºC and 6 ºC depending on the ventilation conditions and other factors (López et al., 2012a; 2012b; López et al., 2013). Molina-Aiz et al. (2004) studied the vertical profile of air velocity and air temperature in an Almería-type greenhouse, where a maximum temperature difference of 14.5 ºC was measured when low wind speed. Granados et al. (2016) measured the soil and air temperature profiles for different solarisation strategies, with a maximum difference of 9.1 ºC, between 0.2 m and 2.0 m height, of the mean air temperature at 2 p.m. during January to March. In view of these results, it seems logical to think that a typical greenhouse could presents conditions of heterogeneity according to ISO 7726 (1998). Since previous works focus on the climatic conditions involved in crop growth, further specific studies are needed to evaluate the spatial variation of the climatic parameters concerning the assessment of the heat stress.

To meet this need, the implementation of a multi-point measurement system inside a greenhouse would provide relevant information on the temporal and spatial variation of the climatic parameters related to the heat stress in the working environment. This is exactly the area where the main contributions of the present work focus, namely:

- To the best knowledge of the authors, this work is the first one approaching the study of climatic parameters of a greenhouse monitoring multiple climatic variables simultaneously at different locations inside the greenhouse and at the three heights specified by ISO 7726 (which correspond to ankle, abdomen, and head).

- To that aims, we design and develop a custom measurement station, with sensors replicated at three different heights, and a distributed communication network to collect all the data in real time.
- Finally, based on data collected with our sensor network, we assess the heterogeneity condition for a typical greenhouse according to the ISO 7726.

Numerous authors have proposed the use of networks of different sensors to monitor agricultural environments, whether in the field or in greenhouses (Ruiz-García et al., 2009). The most effective communication technologies and protocols for these types of applications have been: Bluetooth (Dursch et al., 2004), XBee (Baronti et al., 2007), WiFi (Anastasi et al., 2009), and RFID (Hamrita and Hoffacker, 2005).

Despite that all these technologies have been used in different projects, today there is sufficient experience to evaluate the advantages and disadvantages of each of them. Bluetooth covers an extremely limited connection distance, offering little flexibility in the format and topology of the communications when there is more than two nodes in the system. WiFi, regulated by the standard IEEE 802.11 (IEEE, 2012), is the most widely used protocol in offices, homes, and smartphones, although its energy consumption is higher than in other alternatives, and therefore it is not ideal for data-acquisition applications that need to function uninterrupted. RFID uses a much lower radio frequency than do other technologies (in the MHz instead of the GHz range), permitting the feeding of even small devices by radio waves, without the need of wires. Its range, however, is seriously limited (hardly 1-2 m), and therefore its use in applications of data application over broad surface areas does not prove optimal, either. The protocol XBee, designed specifically for multiple-node networks, permits the connection in a diversity of topologies, maintaining energy consumption to a minimum. Thus, it seems ideal for applications in agriculture under plastic (Daronti et al., 2004;

Anastasi et al., 2009) and more specifically its variant regulated by the IEEE standard 802.15.4 (IEEE, 2009).

The rest of this work is structured as follows. Section 2 describes the basic climatic parameters and defines the limits of homogeneous environments according to ISO 7726 (1998). Next, the grid of measurement stations distributed inside the greenhouse, the characteristics of the measuring instruments and the network architecture are detailed. Section 3 presents the results of the work, mainly focused on the evaluation of the heterogeneity conditions. Section 4 includes the discussion of the obtained results. Finally, the conclusions drawn from this work are summarized in the last section.

## 2. Material and Methods

### 2.1. Climatic parameters

ISO 7726 (1998) defines as basic climatic parameters the air temperature ($t_a$), the mean radiant temperature ($\bar{t_r}$), the air velocity ($v_a$), and the air humidity (expressed by the partial vapour pressure, $P_a$). The mean radiant temperature ($\bar{t_r}$) can be calculated from the black globe temperature ($t_g$) as

$$\bar{t_r} = \left[(t_g + 273)^4 + 0.4 \cdot 10^8 \cdot |t_g - t_a|^{1/4} \cdot (t_g - t_a)\right]^{1/4} - 273 \qquad (1)$$

when natural convection and a normalized black globe of 15 cm in diameter is used (ISO 7726, 1998). In case of forced convection, the following equation can be used:

$$\bar{t_r} = \left[(t_g + 273)^4 + 2.5 \cdot 10^8 \cdot v_a^{0.6} \cdot (t_g - t_a)\right]^{1/4} - 273 \qquad (2)$$

The selection between natural and force-convection equations require the calculation of the coefficient of thermal transmission ($h_{cg}$), which is defined, for natural convection as

$$h_{cg} = 1.4 \cdot \left(\frac{t_g - t_a}{D}\right)^{1/4} \tag{3}$$

and, for forced convection, as:

$$h_{cg} = 6.3 \cdot \left(\frac{v_a^{0.6}}{D^{0.4}}\right) \tag{4}$$

where $D$ is the diameter of the black globe. Then, the larger of the two coefficients determined by the Eqs. 3 and 4 determines the type of convection to be used (natural of forced).

From the basic parameters other climatic parameters can be derived and the comfort and heat-stress indices can be evaluated; as the operative temperature ($t_o$), required for PMV index calculation (ISO 7730, 2005), or the natural wet-bulb temperature ($t_{nw}$), required for the WBGT index equation (D'Ambrosio-Alfano et al., 2004).

Also, these basic parameters are used in ISO 7726 (1998) to define the limits of homogeneous environments. Regarding the spatial variation of the physical parameters within the workplace, ISO 7726 (1998) establishes some limits for the basic parameters where, if the variation of the basic parameters is within a given range, the environment can be considered homogeneous, whereupon only one measurement point is required. Otherwise, if the variation of the climatic parameters exceeds the previous range, the work space must be evaluated as a heterogeneous environment. Table 1 summarizes, from the mean value, the maximum admissible deviation of the basic parameters measured throughout the workplace to be considered homogeneous. The homogeneity of the environment is evaluated for each basic parameter individually —that is, the space can be homogeneous in air temperature but heterogeneous in radiant temperature.

Table 1. Maximum deviations* of the climatic parameters for homogeneous environments (ISO 7726, 1998).

| | |
|---|---|
| Air temperature ($t_a$) | ± 2.0ºC (0ºC < $t_a$ < 50ºC) |
| Mean radiant temperature ($\bar{t_r}$) | ± 10.0ºC (0ºC < $\bar{t_r}$ < 50ºC) |
| Air velocity ($v_a$) | ±(0.3+0.15· $v_a$) m s$^{-1}$ |
| Partial vapour pressure ($P_a$) | ± 0.45 kPa |

* For thermal environments where heat stress is possible (type S).

The spatial heterogeneity of the environment must be evaluated vertically and horizontally. In vertical, ISO 7726 (1998) sets three heights where the physical quantities are to be measured: ankle, abdomen, and head. The specific value of these heights can be selected to fit the characteristics of each population. Weighting factors of 1, 2, and 1, for the ankle, abdomen, and head, respectively, are used to calculate the mean value of these three measurements. In this work, a measurement station has been designed to include sensors at three heights (see Section 2.3).

To gather information on the variation of the climatic parameters in the horizontal direction, a grid of measurement stations was distributed throughout the greenhouse. Each station collected data of the air temperature ($t_a$), black globe temperature ($t_g$), air velocity ($v_a$) and relative humidity (RH).

Beyond the scope of the thermal-environment assessment rules, each measurement station was equipped with an ultraviolet (UV) sensor in order to measure de UV incidence inside the greenhouse. The guideline ISO 17166 (1999) describes the formulation of the worldwide solar ultraviolet index (UVI) based on the reference action spectrum of the International Commission on Illumination (CIE) for the erythema induced by UV on human skin. This UV radiation goes from 100 to 400 nm, divisible into three well-defined intervals: UVA (315-400 nm), UVB (280-315 nm) and UVC

(100-280 nm) radiation. The human work space is reached by UVA wavelengths and a very low percentage of UVB (WHO, 2002).

UVI measures the intensity of solar UV radiation on the work surface, this being a dimensionless index classified worldwide by a colour code. This is determined as a function of a constant of 40 m²·W⁻¹ ($k_{er}$), of spectral solar radiation at a wavelength λ ($E_\lambda$ in W·m⁻²·nm⁻¹), of the reference action spectrum for the erythema ($S_{er}(\lambda)$] and of the wavelength differential used in the integration according to Eq. (6):

$$\text{UVI} = k_{er} \int_{250\ nm}^{400\ nm} E(\lambda) S_{er}(\lambda) d(\lambda) \tag{6}$$

UVI can be calculated using this Eq. (6) or using a broadband detector calibrated and properly programmed to give the UVI values directly (WHO, 2002). This last method has been used in this work.

**2.2. Monitored greenhouse**

In heterogeneous environments, physical quantities must be measured at the different locations where the subject may be located. Partial results should be considered to determine the mean values of the required quantities for assessing heat stress. As a presumable heterogeneous environment, a grid of sensors were installed in an Almería-type greenhouse, located at 15 km east of the city of Almería, Spain (36º52' N – 2º17' W, 98 m a.s.l.). The greenhouse had an area of around one ha (32 m x 32 m) and a mean height of 3.7 m (Figure 2). The supports were of steel and the cover was a three-layer polyethylene plastic film of 200 μm and 81% visible light transmittance. Ventilation was provided through roof vents and lateral windows. The main characteristics of the greenhouse are listed in Table 2. Inside the greenhouse, a main lane was intersected by the secondary lanes flanked by the rows of tomato plants. A grid of 12 measurement stations were distributed along the greenhouse in order to record representative data of the entire workspace (Figure 2 and Figure 3). In addition, another measurement station was placed outside the greenhouse to measure the outdoor climatic conditions. Measurement started on 7 September 2016, before transplanting the tomato plants, and was planned to end in September 2017, in order to collect data over a complete year. Measurement data is meant to provide relevant information on the variation of the climatic parameters over time and space in the greenhouse.

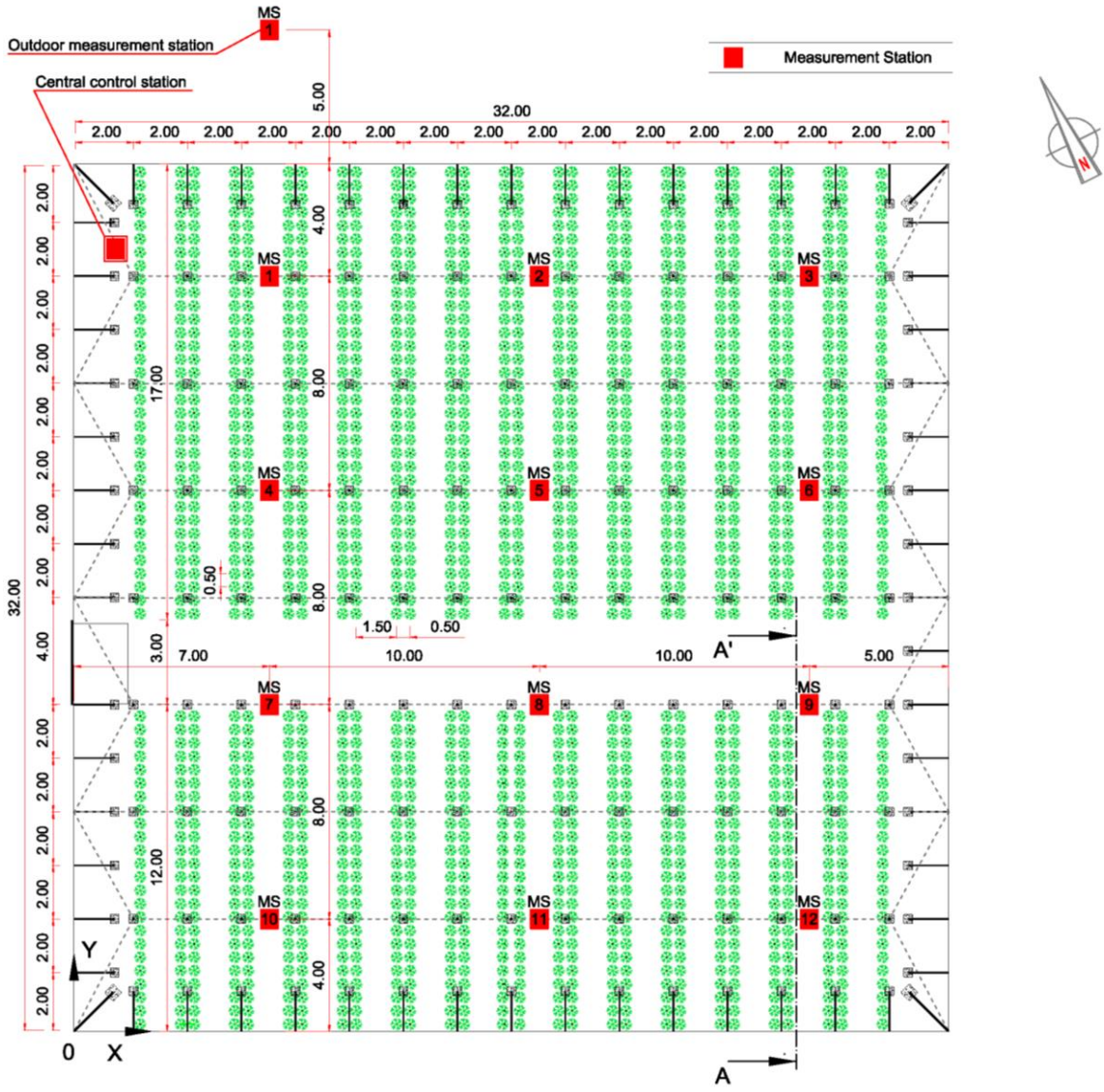

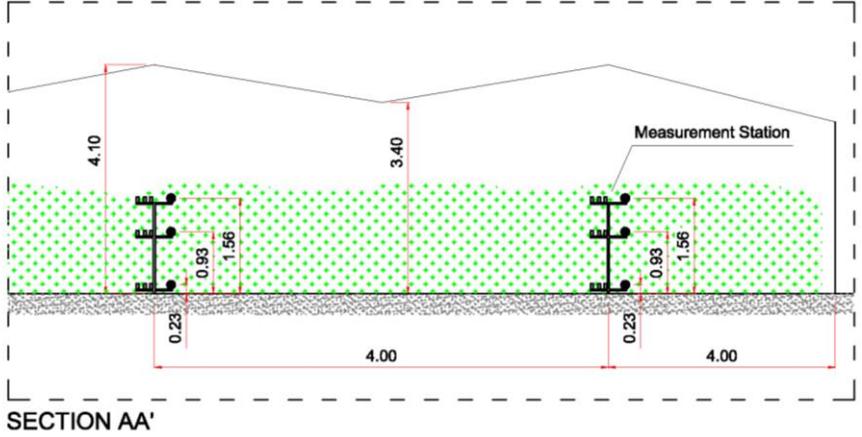

Figure 2. Greenhouse views and locations of measurement stations.

Table 2. Characteristics of the greenhouse.

| | |
|---|---|
| Size | 32 m x 32 m; h = 3.40÷4.10 m |
| Cover | Three-layer polyethylene film; 200 μm thickness; 81% visible light transmittance; 29% diffuse light transmittance |
| Openings | Manual lateral windows and roof automatic vents. |
| Shading | White washing applied at 07 Sept. (removed when first rains at 29 Oct.) |
| Floor | Gravel-sand-covered soil |
| Crop | Tomato (drip irrigation); Plant height: 2.1 m |

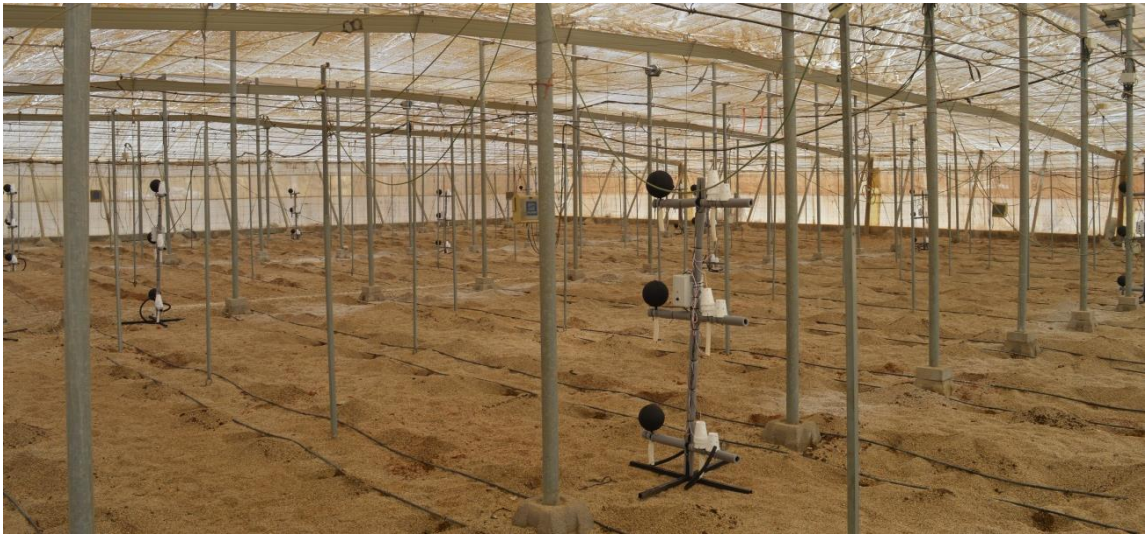

Figure 3. Measurement stations placed inside the greenhouse.

### 2.3. Measurement stations

For compliance with guideline ISO 7726 (1998) in heterogeneous environments, a customized measurement station (MS) was designed. The measurement station included a structural support, several measuring instruments, and an electronic box for signal processing (Figure 4). The support consisted mainly of a vertical post with three horizontal bars adjustable in height. The horizontal bars were fixed at 0.23 m, 0.93 m and 1.56 m from the floor, where these heights were selected from the percentile 50 of the Spanish population (Carmona-Benjumea, 2001).

Each horizontal bar is equipped with four probes for measuring the microclimatic parameters: air temperature ($t_a$), black globe temperature ($t_g$), air velocity ($v_a$), and relative humidity (RH).

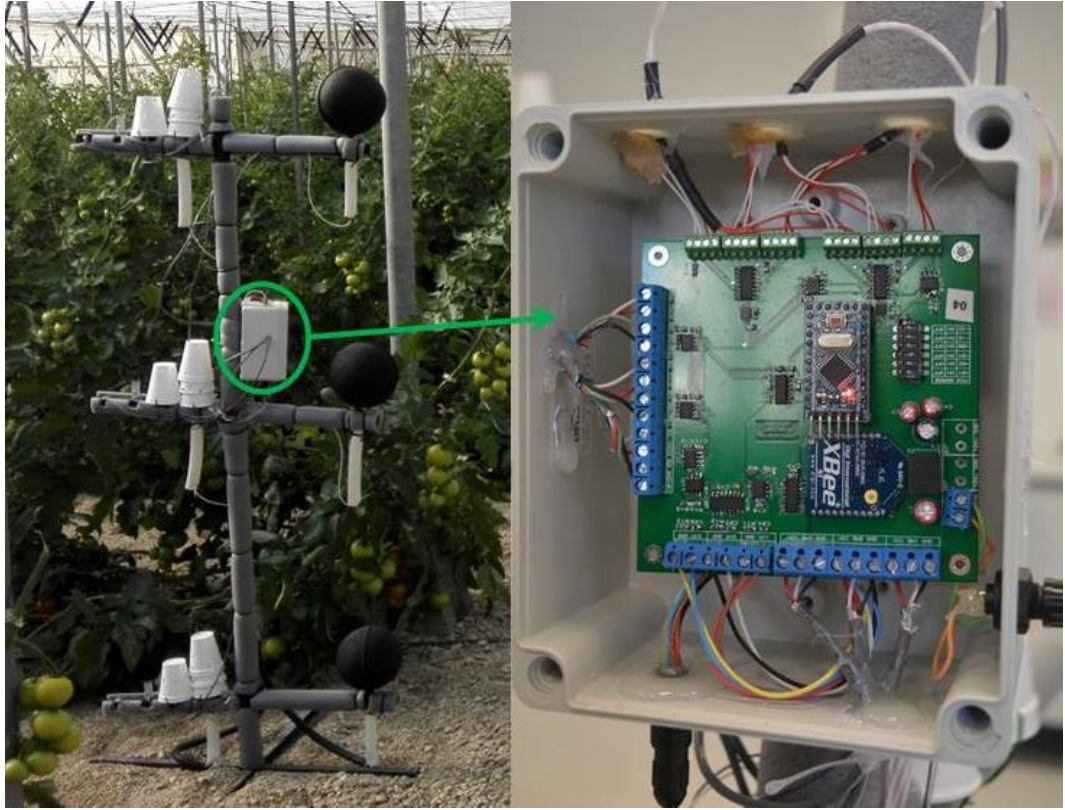

Figure 4. Measurement station and electronic box.

Table 3. Characteristics of the measuring instruments.

| Parameter measured | Manufacturer | Model | Measure range | Accuracy |
|---|---|---|---|---|
| Air temperature ($t_a$) | TC Direct | 515-725 | -15ºC to 250ºC | ± 0.06ºC (at 0ºC) |
| Globe temp. ($t_g$) | TC Direct | 515-725 | -15ºC to 250ºC | ± 0.06ºC (at 0ºC) |
| Air velocity ($v_a$) | Modern Device | Rev. C | 0 to 20 m·s$^{-1}$ | ± 10%, 5 cm·s$^{-1}$ |
| Relative humidity (HR) | Silicon Labs | Si7021-A20 | 0 to 100% HR | ± 3% RH |
| Ultraviolet radiation (UVI) | Silicon Labs | Si1145/46/47 | 0-15 | ± 1 |

The characteristics of each probe, selected to comply with the requirements of the ISO 7726 (1998), are listed in Table 3. For the measurement of the black globe temperature ($t_g$), a Pt100 probe was placed inside a black globe made of brass (66% Cu, 34% Zn), 15 cm in diameter and 0.5 mm thick. Also, Pt100 probes were used to measure the air temperature ($t_a$), where these probes were covered through white ventilated protections to avoid the direct solar radiation. The humidity sensor, with a measurement range of 0% to 100%, was similarly covered for protection against the condensed water dripping from the plastic-film roof. A hot-wire anemometric sensor with a measure limit of 20 m·s$^{-1}$ was used to measure the air velocity ($v_a$), which is far below this limit. The limitation of using this low-cost hot-wire anemometric sensor is its need to be calibrated. For the calibration, a 2-axis ultrasonic wind sensor (Gill WindSonic) was also installed for a period of two weeks. The hot-wire anemometric sensor was not calibrated in the present work, as this will be undertaken in a future work.

In addition to the four sensors described, an ultraviolet radiation sensor was placed in the upper bar of each measurement station. The aim of this sensor installation was to measure the incidence of UV radiation inside the greenhouse, the absorption of UV radiation due to the plastic film of the greenhouse cover, and the degeneration of this film property over its service life (3 years). Then, beyond the first year, when all sensors will be operating, UV sensors will continue working for two more years to evaluate the UV incidence inside de greenhouse.

All sensors were wired to an electronic box where the signals measured were converted and processed. The electronic box was cable powered from an uninterruptible power supply.

**2.4. Network architecture**

As shown in Figure 5, there are four elements: (1) measurement stations, distributed throughout the greenhouse being monitored; (2) the central control station, installed in a box within the greenhouse itself; (3) a server, installed in the Data Processing Centre of the University of Almería (CPD-UAL) and, optionally, (4) remote operators.

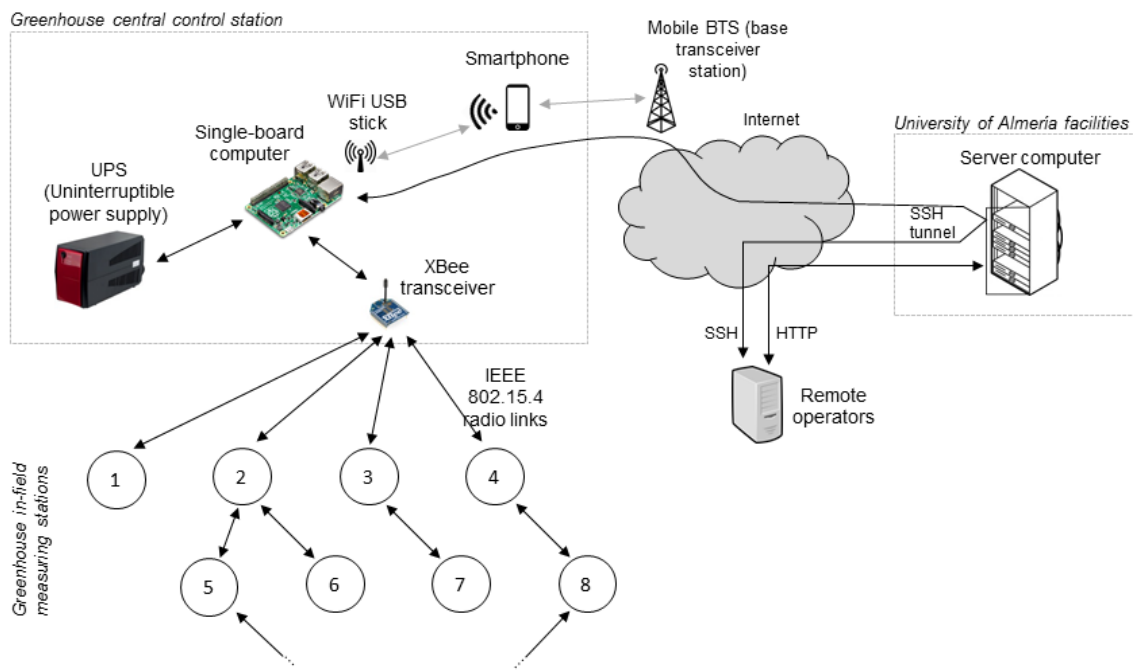

Figure 5. General view of the architecture of the system designed.

The core of the system consists of measurement stations and a central station, all installed locally in the greenhouse, and having totally autonomous operating capacity. The outside elements (server and operators) are added to improve the usability of the system on indicating its state in real time from any place on the planet with access to Internet. The server installed in the Data Processing Centre of the University of Almería (CPD-UAL) was also used to store backups of all the data compiled daily by the central station of the greenhouse, so long as there was wireless connection and cover in the area.

The measurement stations communicated to a central node and also among them when necessary, using point-to-point low-power (<10 mW) IEEE 802.15.4 radio links in the 2.4 GHz band. The firmware of the measurement stations and the software of the central node (both developed in C++ and compiled for AVR8 and ARM architectures, respectively) were designed such that a reconfigurable routing table could be used to define the desired communication routes for the exchange of data packets between the central node and the base stations. In this way, the maximum range of the radio signal was expanded using stations as intermediary relays at the radio packet level, which proved to be a valuable property given the large radio attenuation that typically exist in environments with high density of plants.

Every 20 seconds, the central control station sent a package to all the measurement stations, indicating that they should sample the data from all the sensors. In this way, it was ensured that all the stations sampled the data at the same time. After a few seconds, the time required to acquire all the analogical signals by precision ADC, the central node began a round of polling to require each station to send the values measured. All the data were stored in structure of directories and files that facilitate the daily backups. As long as there was Internet connection, by means of a Smartphone 4G, the central node sent the daily data to a server located in the CPD-UAL. The operators (researchers of this project) could gain access to these data on the server by a simple web interface (HTTP), or by safe Secure Shell (SSH) links.

In addition, the system was configured to permit connections from each place equipped with Internet access to the central computer of the greenhouse. This was necessary because as the central computer was behind the NAT (Network Address Translation) of the mobile operator, which blocks direct connections from outside networks. The solution we adopted was to create a reverse SSH connection from the central node of the greenhouse to the server in the CPD-UAL, established 24 h per day in case of being required by an outside operator.

Regarding memory and storage requirements of our system, our embedded microcontroller-based platform has a minimal memory footprint: only 17 KiB of program memory (flash memory) and 0.8 KiB of data memory (RAM). Only a few KiB are also required for the ARM-based program in the central node. Storage of all collected data typically requires 16 MiB per day if stored uncompressed.

The inner architecture of each measurement station is diagrammed in Figure 6 (also see its implementation in Figure 4). The system was based on a low-cost, low-energy-consumption AVR8 (Atmel) microcontroller. The IEEE 802.15.4 radio network was connected by a software layer that provided access to the API of a XBee S1module physically connected by a standard UART.

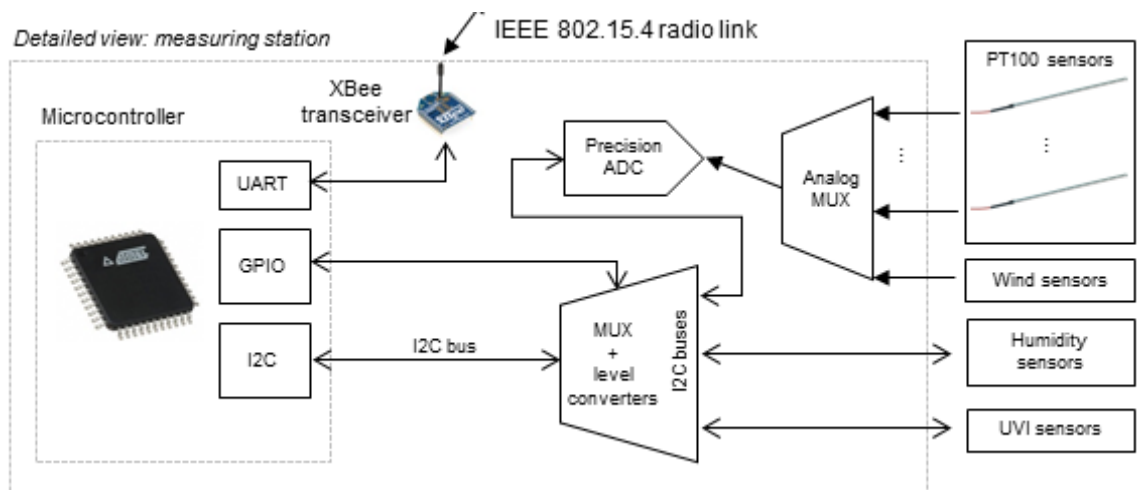

Figure 6. Electronic modules comprising each of the measurement stations.

The rest of the sensors were accessible by a I2C bus, which was multiplexed so that it could read a multitude of devices connected simultaneously in the same bus. Voltage converters were included to make the devices compatible with both, 5 V and 3.3 V

CMOS logic levels. Finally, the analog sensors (PT100 temperature probes, "hot wire" wind sensors) were acquired, after the corresponding analog stage, by a high-precision ADC (Analogical-to-digital converter) that provides resolution on the order of 1 µV.

## 2.5. Comparison of proposed network to other works

After describing the measuring instruments of our stations in section 2.3 and the implemented communication and control network in section 2.4, we proceed to discuss how our proposal compares to other sensor networks proposed in the existing literature. Please, refer to Table 4 for a summary of the following discussion.

Regarding the application of the works, we find that most of them focus on crop growth (Vox et al., 2014; López et al., 2009; Srbinovska et al., 2015; Konstantinos et al., 2014; Balendonck et al., 2014; López et al., 2012a), whereas another large group also aims at assessing the thermal stress of the measured environments (Pérez-Alonso et al., 2011; Cecchini et al., 2010; Callejon-Ferre et al., 2011a; Marucci et al., 2012). Our proposal is the only one designed to determine whether a given space is *heterogeneous*, in the sense we introduced in section 1, all along the three dimensions of the space. Only a few works, like (Ferentinos et al., 2017; Balendonck et al., 2014), address the *horizontal* heterogeneity of climatic variables. This unique feature of our sensing stations can be seen in Table 4, where it is clear that no other work proposed placing sensors at three different heights. To evaluate the horizontal heterogeneity in our work, a total of 12 measurement stations have been equally distributed inside the greenhouse. This number of stations is in the order of magnitude of the largest numbers used in previous works for greenhouses of similar dimensions, which vary between 5 and 12 stations.

Of all previous works proposing a Wireless Sensor Network (WSN), most of them employ IEEE 802.15.4 plus ZigBee as communication protocol between the sensing nodes and other nodes or a central hub. Indeed, the low-energy cost of this protocol makes it an obvious choice for a WSN, hence we also used it in the present study. Still, some works place them apart of this hegemony. For example, (Hamrita and Hoffacker,

2005) employed RFID tags for short range measuring, and (Anastasi et al., 2009) used IEEE 802.11b (best known as "WiFi") to communicate between nodes in the WSN.

Network topology is another central design parameter in any WSN. All proposed works, including ours, include a central node in which all measurements are centralized as they are gathered from a number of smaller and simpler sensing stations or nodes. As can be seen in Table 4, roughly half of the WNS use a star topology (a central node is connected to all sensing nodes), while in the other half communication happens following a tree (i.e. a network without loops) between the central node and the sensing stations. ZigBee allows both topologies to be straightforwardly configured via a pre-programming of configuration parameters in the communication modules. The advantage of the tree topology is an extended area of coverage for deploying sensors, since the ultimate communication range limitation is the peer-to-peer maximum range. Our architecture exhibits another unique feature in this area, since peer-to-peer communication can happen in a mesh (all-to-all) topology, which is exploited in our firmware and software to generate dynamic routing tables such that communication to distant stations always follow a tree, but whose components can be changed on-the-fly even from a remote-control station via Internet.

Regarding the communication between the central nodes and the remote-control station, we find different alternatives in the literature. Many works simply do not consider such a remote control and store all measured variables locally without any obvious way to access the data online. In turn, others like (López et al., 2009; Srbinovska et al., 2015) use radio modems with different digital modulations to establish a link with a computer external to the measured environment. Clearly, this limits the attainable range to a maximum of a few kilometres. Others, such as (Vox et al., 2014), use the same approach as our proposal and employ the public mobile phone network to enable the remote access to the WSN.

Finally, the accuracy of each measuring instrument in related works is also shown in Table 4. Regarding temperature sensing, it is remarkable that our stations are the only ones equipped with the high-accuracy 1/10 DIN Pt-100 sensors. The rest of sensor accuracies are relatively similar among all works.

Table 4. Summary of related works and the characteristics of their proposed sensor networks.

| Work | Application | Network topology | Communications technology | Number of stations / heights | ISO 7726 application | Accuracy of measurement instruments ||||||
|---|---|---|---|---|---|---|---|---|---|---|---|
| | | | | | | Air temperature | Damp bulb temperature | Globe temperature | Relative humidity | Air velocity | Others |
| Hamrita & Hoffacker (2005) | Crop Growth (soil properties). | None | RFID | 1 / 1 | No | | | | | | Soil temperature TMP04 ±1 °C |
| Anastasi et al. (2009) | Wine production quality. | Tree | IEEE 802.15.4 / ZigBee IEEE 802.11b ("WiFi") | 12 / 1 | No | SHT11 ±0.5 °C | | | ±3.5% RH | MTX VO 009 1,5 m·s$^{-1}$ | PAR light |
| López et al. (2009) | Crop Growth. | Tree | IEEE 802.15.4 / ZigBee Radiomodem FHSS, FSK | 10 / 1 | No | SHT71 ±0.4 °C @25 °C | | | SHT71 ±3.5% RH | | Water electrical conductivity and water temperature. Soil moisture, conductivity, salinity and temperature |
| Cecchini et al. (2010) | Thermal stress. | None | Direct sensor connections | 1 / 1 | Yes | Pt-100 1/3 DIN | Pt-100 1/3DIN | Pt-100 1/3DIN | ±2% RH | Hot wire 0.04 m·s$^{-1}$ | |
| Pérez-Alonso et al. (2011) | Thermal stress. Greenhouse-construction. | None | Direct sensor connections | 1 / 1 | No | Pt-100 1/3 DIN | Pt-100 Class A | Pt-100 1/3DIN | ±2.5% RH | Hot-wire 0.1 m·s$^{-1}$ | Pyranometer- Solar radiation |
| Callejon-Ferre et al. (2011a) | Thermal stress. | None | Direct sensor connections | 1 / 1 | No | Pt-100 1/3 DIN | Pt-100 Class A | | ±2.5% RH | Hot-wire 0.1 m·s$^{-1}$ | |
| Marucci et al. (2012) | Thermal stress. | None | Direct sensor connections | 1 / 1 | Yes | Pt-100 1/3 DIN | Pt-100 1/3DIN | Pt-100 1/3DIN | ±2% RH | Hot wire 0.04 m·s$^{-1}$ | |
| López et al. (2012a) | Crop Growth. | None | Direct sensor connections | 12 / 2 | No | ±0.026 °C | | | | Sonic anemo. (3D) ±0.04 m·s$^{-1}$ | |
| Vox et al. (2014) | Crop Growth. | Tree | IEEE 802.15.4 / ZigBee Mobile phone network (GPRS) | 6 / 1 | No | SHT75 ±0.5 °C | | | SHT75 ±2% RH | Potentiomer (rotating paddles) 1.1 m·s$^{-1}$ | Air pressure; Solar radiation |
| Balendonck et al (2014) | Crop Growth. Horizontal climatic heterogeneity. | Star | WiSensys (not specified) | 100 / 1 | No | SHT71 ±0.4 °C @25 °C | | | SHT71 ±5% RH | | |
| Srbinovska et al. (2015) | Crop Growth. | Star | IEEE 802.15.4 / ZigBee Radiomodem FSK | 5 / 1 | No | SHT11 ±0.5 °C | | | SHT11 ±3% RH | | |
| Ferentinos et al (2017) | Crop Growth. Horizontal climatic heterogeneity. | Star | IEEE 802.15.4 / ZigBee | 5 / 1 | No | SHT11 ±0.5 °C | | | SHT11 ±3% RH | | |
| Ours | Ergonomics of the thermal environment. Heterogeneous. | Physical mesh, logical tree | IEEE 802.15.4 / ZigBee Mobile phone network (G4) | 13 / 3 | Yes | Pt-100 1/10 DIN (±0.06 °C) | | Pt-100 1/10 DIN (±0.06 °C) | ±3% RH | Hot wire ±10%, 0.05 ms$^{-1}$ | UVI |

## 3. Results

This section presents climatic data from the measurement stations. Most of the results are focused on confirming the heterogeneity conditions inside the greenhouse. The measurement stations and the network developed successfully provided climatic data at twelve locations and three heights inside the greenhouse. For an example of the data provided by the stations, Figures 7 and 8 show the values of the climatic parameters collected by measurement station nº 1 over a week, from 22 to 28 January 2017.

Despite that it was the winter season, air temperatures of over 25ºC were reached at midday on sunny days (Figure 7a). By contrast, maximum temperatures of around 15ºC were measured on cloudy days (22 and 27 January). It is also relevant the differences in the air temperature between each sensor height, with differences over 5ºC on sunny days between the upper sensor (at 1.56 m) and the lower one (at 0.23 m). Black globe temperature showed the same pattern as the air temperature but, as expected, reached higher values with the presence of solar radiation, while at night it equalled air temperature (Figure 7b). Relative humidity normally reached the value of 100% at night but fell around 60% at noon (Figure 8a). Also, differences in humidity values are appreciable between the three heights registered by the measurement station. In the case of the wind sensors, Figure 8b shows the rough data of voltage once filtered. The air-velocity data require subsequent treatment according to the calibration curve drawn for the sensor. Finally, Figure 8c presents the data from the UV sensor situated on the upper bar of the measuring station, where the sudden spikes shown on the curve may be due to clouding or clearing.

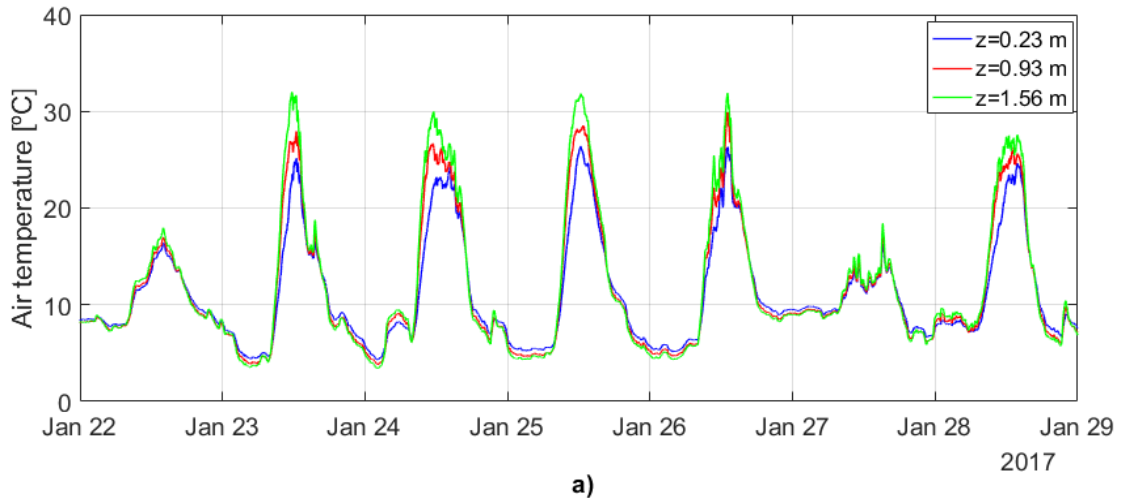

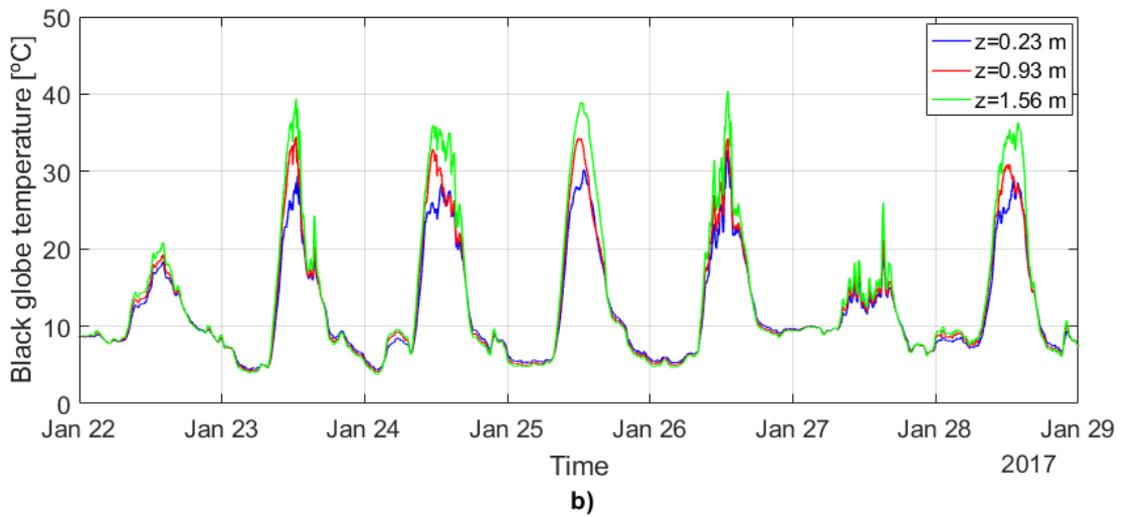

Figure 7. (a) Air temperature and (b) black globe temperature measured by the measurement station #1 from 22 to 28 January 2017.

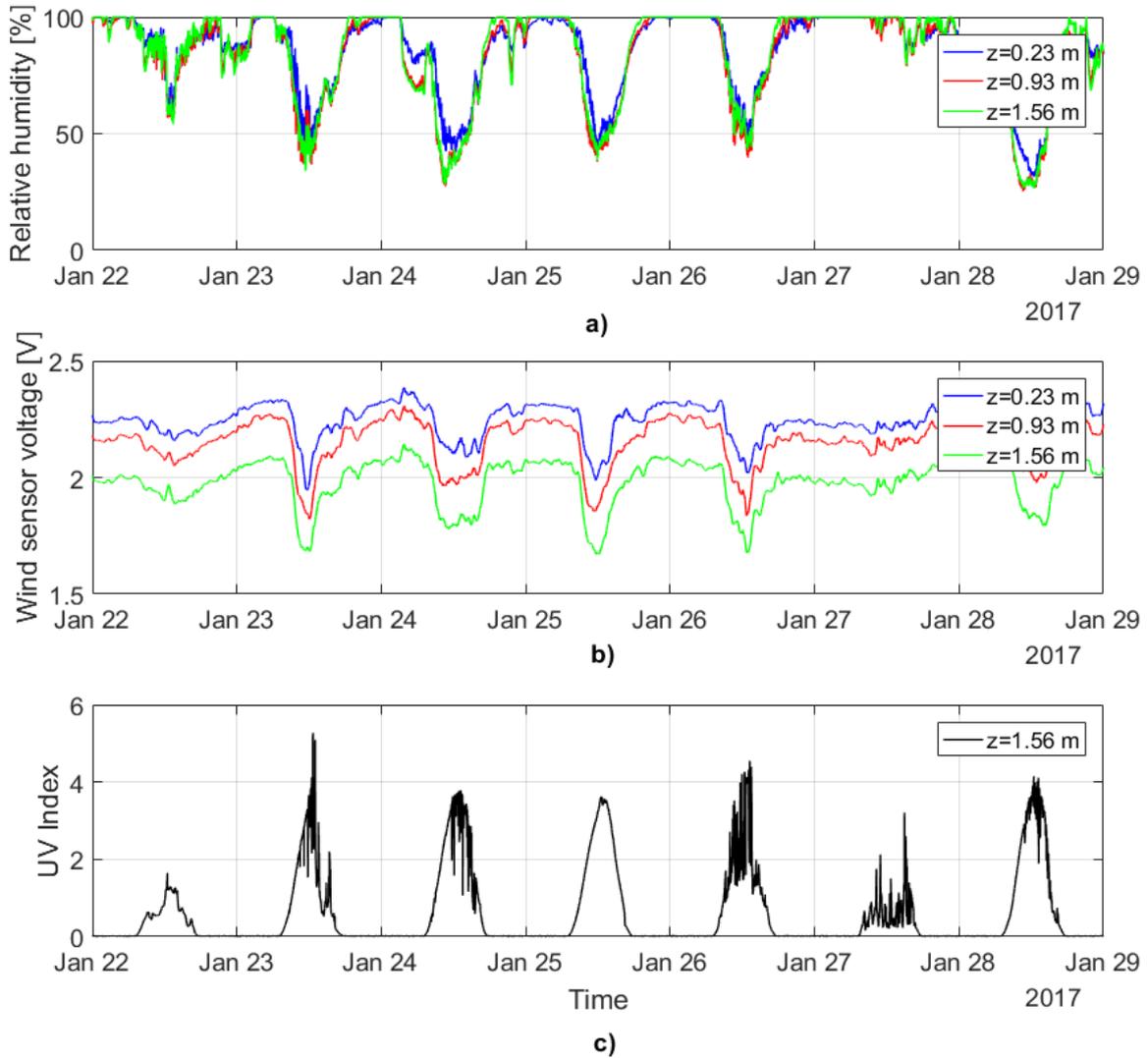

Figure 8. (a) Relative humidity, (b) wind sensor voltage, and (c) ultraviolet radiation measured by the measurement station #1 from 22 to 28 January.

Focusing on the temperature distribution inside the greenhouse, Figure 9 shows the air temperature measured with each individual sensor, grouped according to the three measurement heights, on January 25th. Temperatures at the upper horizontal planes (0.93 m and 1.56 m height) show more uniformity along the greenhouse than in the lower plane (0.23 m height), while the air temperature increase with the height at daylight hours. Global vertical heterogeneity is shown in Figure 10a. Each of the three colour curves represents the mean value of the air temperature at each horizontal plane, i.e. the mean value of each group of twelve curves in Figure 9. The same Figure 10a represents mean air temperature of the greenhouse (black curve with a continuous line), calculated with the weighted mean of the three previous temperatures, and the limits for the space to be considered homogeneous (black curve with a broken line). It can be seen that the greenhouse should be considered a heterogeneous environment between approximately 9:00 UCT and 13:00 UCT, when the air temperature at the upper and lower measurement heights exceed the limits of homogeneity (±2.0ºC; Table 3). The same curves have been drawn for mean radiant temperature (Figure 10b). In this case, the greenhouse can be treated as a homogeneous environment with respect to mean radiant temperature. This pattern of air temperature and mean radiant temperature holds for most days. Over the months of September to January, the mean radiant temperature exceeds the limits of vertical homogeneity in the greenhouse only a few days and for very short periods of time, while the air temperature usually shows periods of heterogeneity of between 2 and 5 hours in the middle of the day. Figure 11 shows the vertical heterogeneity for the air temperature between January 22$^{nd}$ and 28$^{th}$, whereas days 23, 24, 25, 26, and 28 register vertical heterogeneity, and days 22 and 27 can be treated as homogeneous. From Figures 8c and 11, it can be deduced that the heterogeneity conditions tend to occur on sunny days, while cloudy days give rise to homogeneity.

It bears clarifying that radiant temperature is calculated from Eq. 1, assuming conditions of natural convection, which are satisfied only for low air velocities, normally below 0.1 m/s. Such velocities are usually surpassed inside greenhouses, with maximum values registered in previous works of 0.47 m·s$^{-1}$ (López et al., 2012b) and 0.85 m·s$^{-1}$ (Molina-Aiz et al., 2004). Therefore, the results for mean radiant temperature should be considered rough estimates, as more exact values would require considering forced-convection conditions (Eq. 2) once the air velocity is known. Figure 12 shows the mean radiant temperature value for measurement station nº 1, calculated considering natural convection and forced convection for wind velocities up to 1.0 m s$^{-1}$. Then, real mean radiant temperature will be within the area enclosed for these curves.

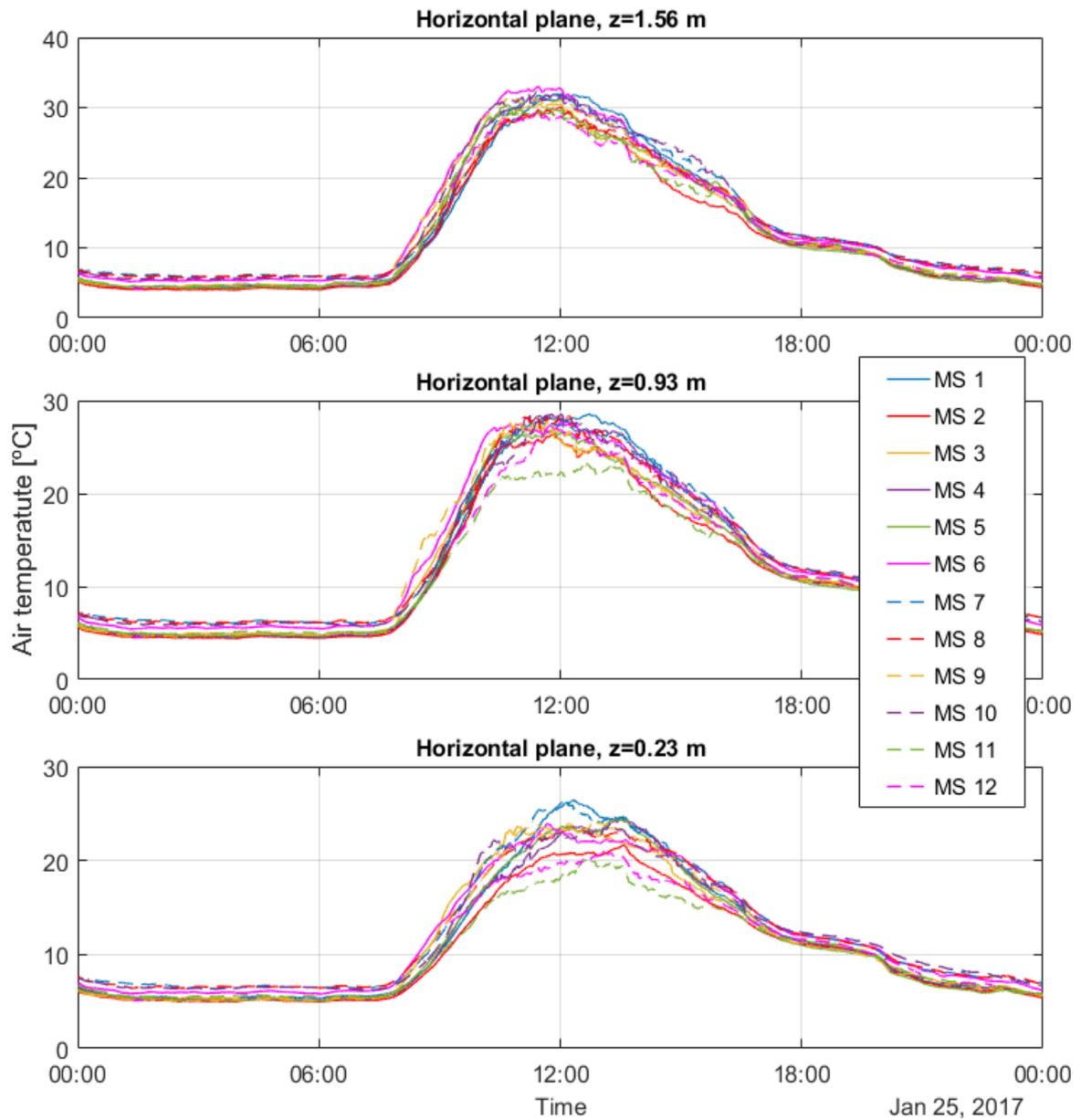

Figure 9. Air temperature measured for each measurement station at 1.56 m, 0.93 m, and 0.23 m height (January 25$^{th}$, 2017).

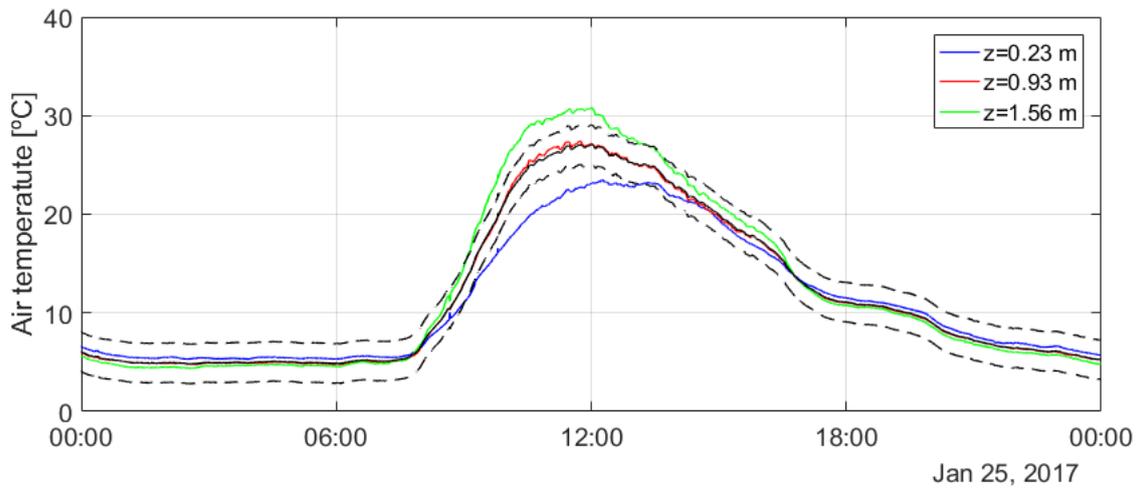

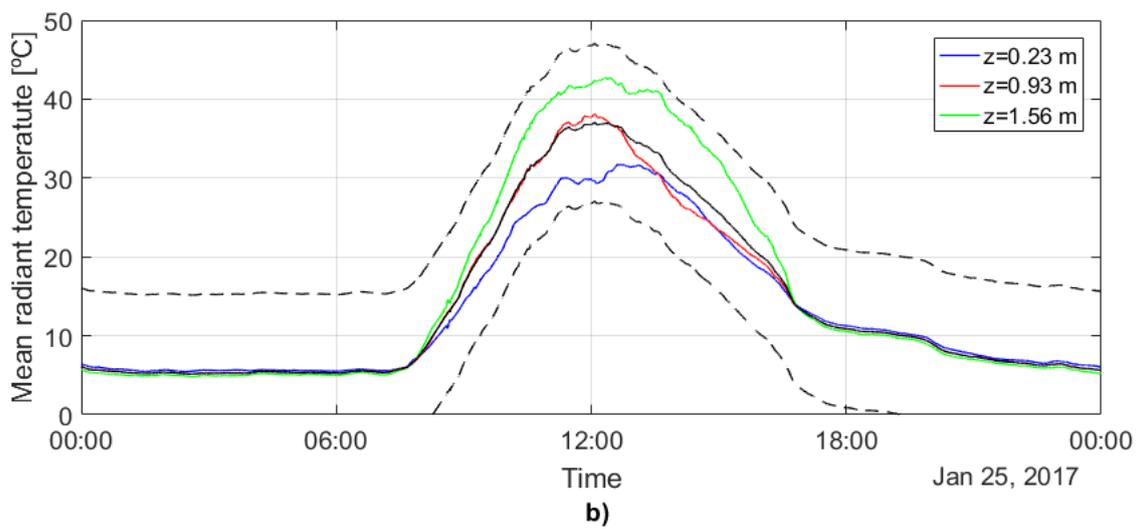

Figure 10. Vertical heterogeneity for (a) air temperature and (b) mean radiant temperature (January 25th, 2017). Continuous black line is the weighted mean value; discontinuous lines are the homogeneity limits.

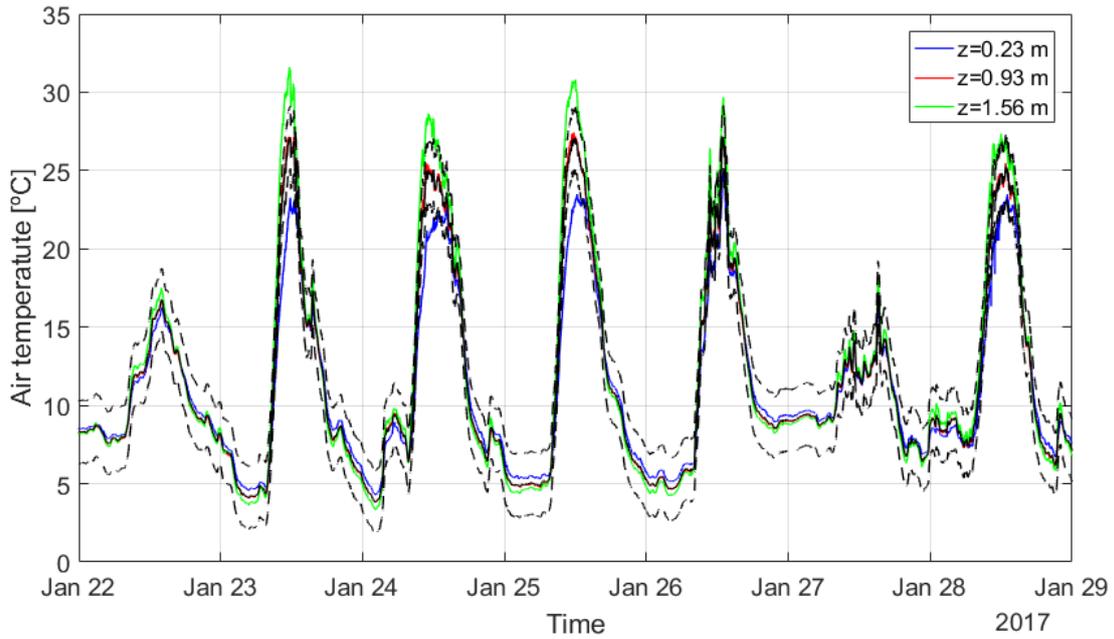

Figure 11. Vertical heterogeneity for air temperature during the days of 22 to 28 January. Continuous black line is the weighted mean value; discontinuous lines are the homogeneity limits.

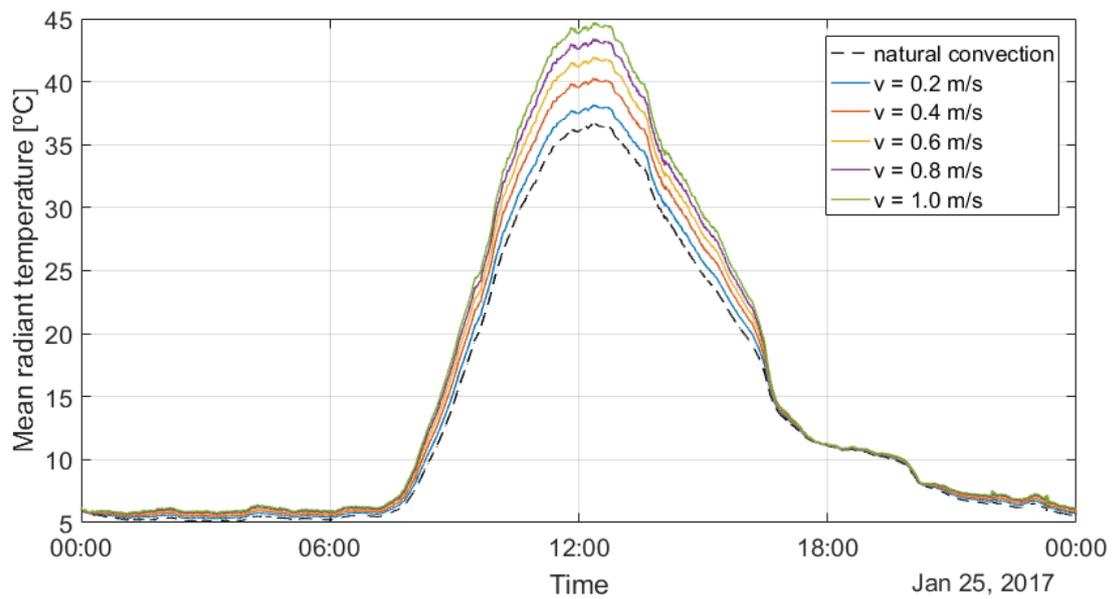

Figure 12. Mean radiant temperature for the case of natural convection and for forced convection at different wind velocities (January 25[th], 2017).

In relation to horizontal heterogeneity, Figures 13 show the results for air temperature and mean radiant temperature (January 25th, 2017). The colour curves show the value for each measurement station, calculated as the weighted mean of the three measurement heights. The mean air temperature and mean radiant temperature of the greenhouse was calculated from the values of the twelve measurement stations (black curve with a continuous line), together with the limits for the consideration of a homogeneous environment (black curve with a broken line). Although a different path was followed, logically the curve for mean air temperature inside the greenhouse is the same in Figures 10a and 13a. As with the results for heterogeneity in vertical, the greenhouse showed horizontal heterogeneity with respect to air temperature, though this usually appeared with one to two hours of lag with respect to the vertical heterogeneity. No horizontal heterogeneity was found with respect to mean radiant temperature during any of the days of the study period. Figure 13a shows that the measurement station nº 11 was under the lower limit for heterogeneity between 10:00 UTC and 17:00 UTC. Another three measurement stations (nº 1, nº 5, and nº 12) also exceeded the limits of homogeneity but during shorter time periods. Figure 14 presents the horizontal heterogeneity for air temperature between January 22$^{nd}$ and 28$^{th}$, showing the conditions of heterogeneity on days 23, 24, 25, 26, and 28 January, the same days on which vertical heterogeneity was registered. The results for horizontal heterogeneity of the greenhouse over the months of September to January showed similar patterns.

These results show that the thermal conditions within the commercial greenhouse were heterogeneous horizontally as well as vertically according with guideline ISO 7726.

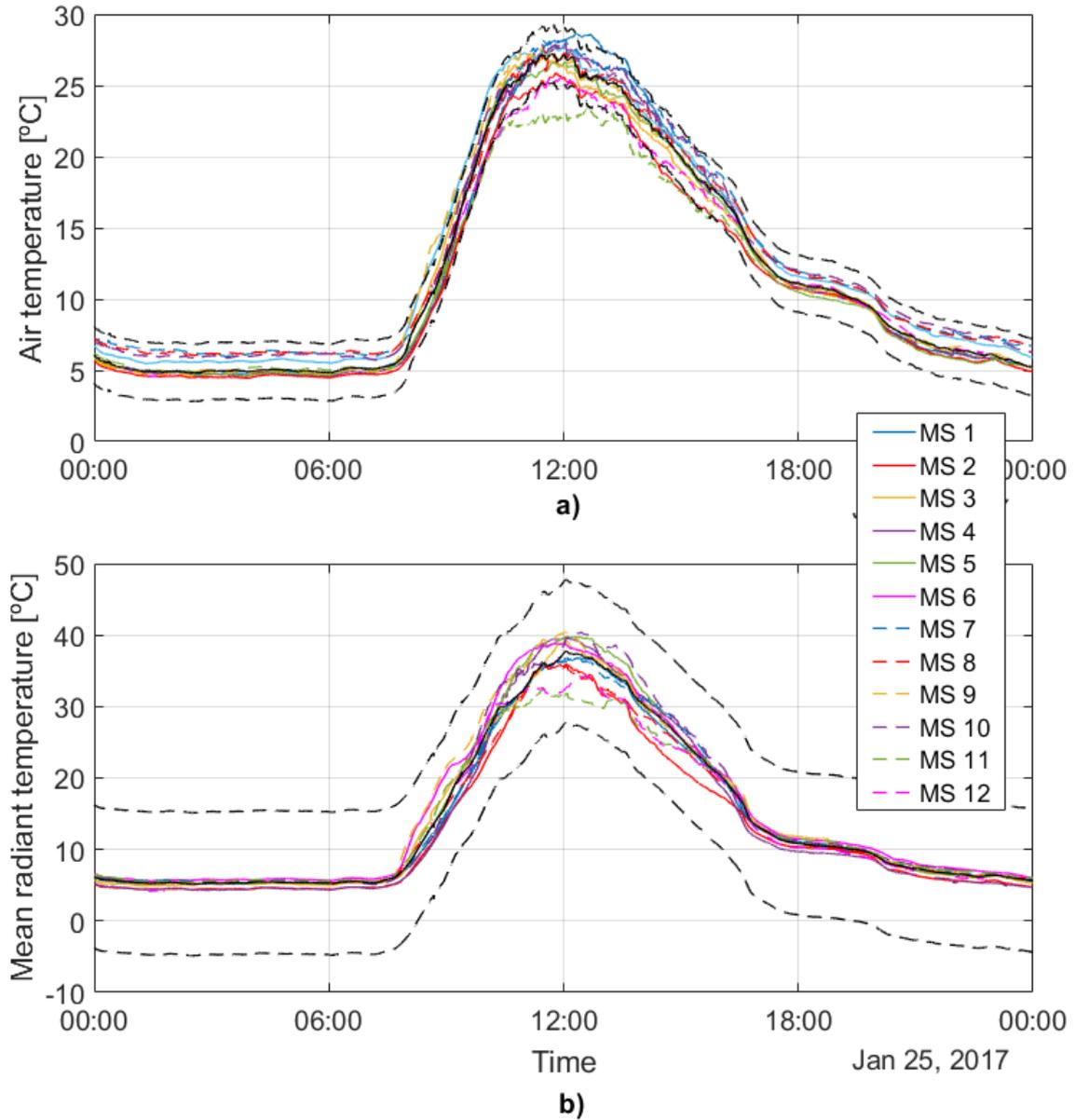

Figure 13. Horizontal heterogeneity for (a) air temperature and (b) mean radiant temperature (January 25th, 2017). Continuous black line is the mean value of the 12 measurement stations; discontinuous lines are the homogeneity limits.

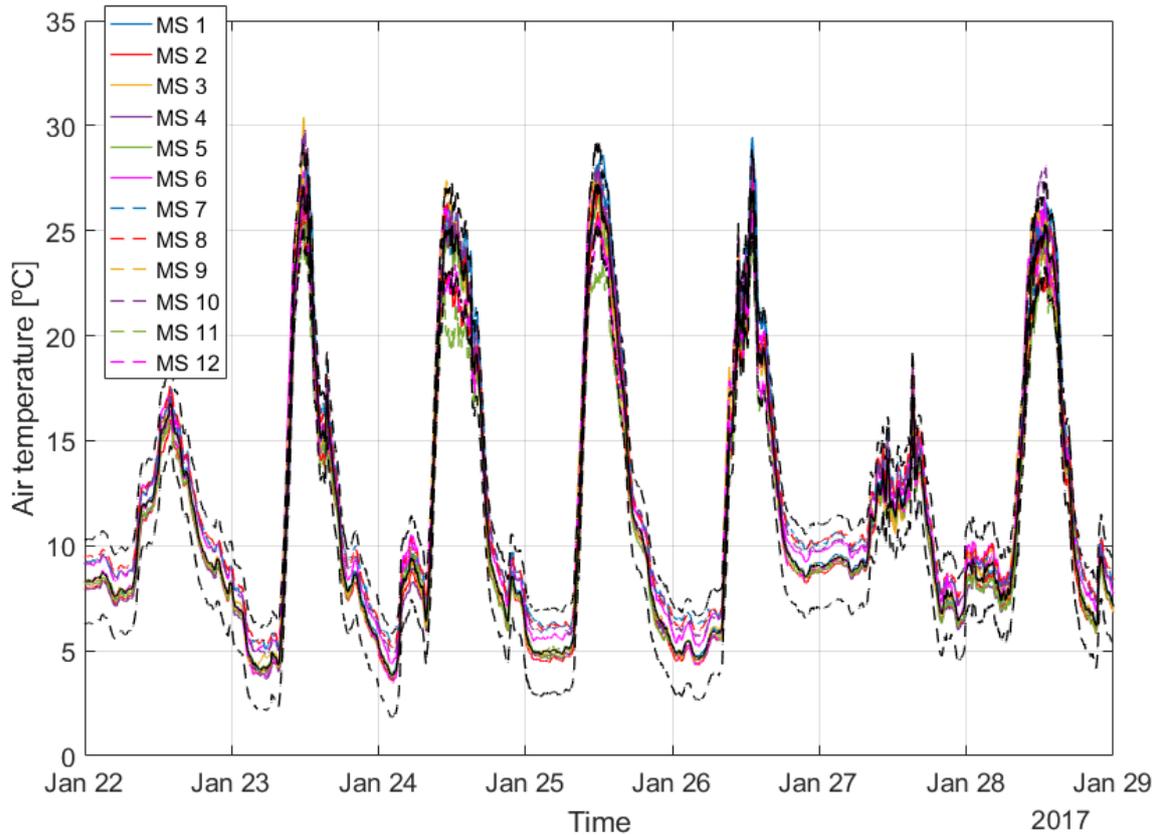

Figure 14. Horizontal heterogeneity for air temperature since January 22nd and the 28th of 2017. Continuous black line is the mean value of the 12 measurement stations; discontinuous lines are the homogeneity limits.

In the present study, a video is included to show in 3D the time course of the air temperature inside the greenhouse (Figure 15 - link). The video depicts the week January 22nd to 28th of 2017, showing several temperature data in the following manner: on the horizontal plane, represented at height zero, the weighted mean air temperature in vertical is shown in a scale of colours. The three vertical planes that contain each of the three rows of the measurement stations show, in a scale of colours, the vertical distribution of air temperature in these planes. In addition, over each measurement station, three markers (blue points) are displaced vertically in a proportional way to the mean temperature for each of the three air temperature sensors. Finally, at the button right of the image, it is indicated the overall maximum and minimum temperatures in every moment among the 24 sensors installed and the partial values at each horizontal plane of 0.23 m, 0.93 m and 1.56 m height.

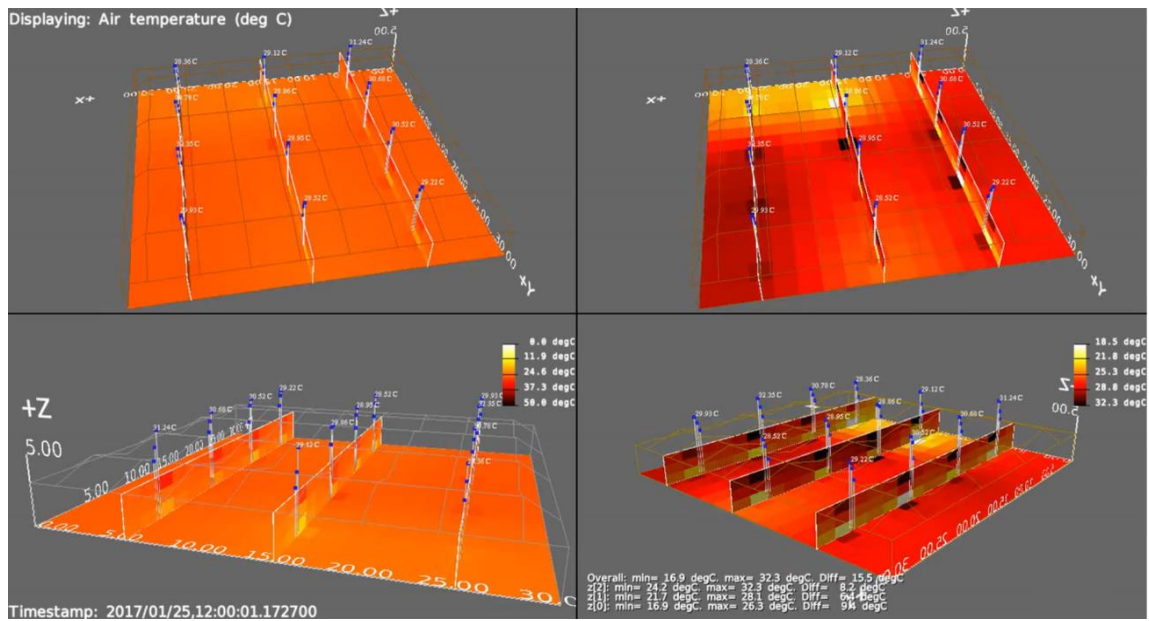

Figure 15. Screenshot from the video showing the air temperature distribution in the greenhouse at 12:00 UTC, January 25th, 2017 (https://vimeo.com/222767975).

**4. Discussion**

The greenhouse where the present research was conducted is the Almería type (Figure 2), being the most representative (c. 90%) (Fernández and Pérez, 2004) of the 30,000 ha of crops under plastic (Cabrera-Sánchez et al., 2016). This explains it choice as being applicable to the immense majority of greenhouse workers in the Almería area.

The measurement stations at three heights, manufactured in the workshop of the Engineering Department of the University of Almería and put into operation in the greenhouse, satisfied the needs of the study. The distribution of the stations inside the greenhouse (Figure 2) provided information on the variation of the environmental parameters in the horizontal and vertical directions and showed the need to consider the greenhouse to be a heterogeneous environment according to the guideline ISO 7726 (1998). The number of measurement stations used and their distribution along the greenhouse has been enough to demonstrate the heterogeneity of the air temperature. A larger number of measurement stations would provide more precise results with respect the exact distribution of the climatic parameters inside the greenhouse, but it seems unnecessary considering the objectives of this works.

For the study period, Figures 11 and 14 indicate that the days with vertical heterogeneity in air temperature also displayed horizontal heterogeneity. The heterogeneity was more accentuated on sunny days while on cloudy days more homogeneous temperatures were registered. During the days of 22 to 28 January, on sunny days, was found maximum differences in mean air temperature in vertical direction (from 0.23 m to 1.56 m) close to 8 ºC at midday (Figure 11). These results are similar to those obtained by Granados et al. (Granados et al., 2016) in a greenhouse with the same type of sand-covered soil, with differences of 9.1 ºC at 2 p.m. (January to March), between 0.2 m and 2.0 m height. In contrast to sunny days, on cloudy days shorter difference in air temperature has been found. Thus, on 22 and 27 January, differences in mean air temperature in vertical direction of around 2 ºC were measured (Figure 11).

In horizontal direction, the air temperature differences were less pronounced than in vertical direction, but also within the consideration of heterogeneous environment. Maximum air temperature differences between 6 ºC (sunny days) and 2 ºC (cloudy days) has been measured (Figure 14). Similar result was found in a multispan greenhouse in a horizontal plane at 1.75 m height (López et al., 2012a).

In addition to these air temperature differences evaluated as the mean values in vertical direction or in horizontal direction, larger differences can be found between the hottest and the coldest individual sensor. For example, a contrast of 14.4 ºC was measured at 12:00 UTC between the lower air temperature sensor at measure station nº 11 (18.3 ºC) and the upper one at measure station nº 6 (32.7 ºC), see Figure 9.

It is necessary to note that the climatic conditions inside a greenhouse will depend on several factors, as the constructive type of greenhouse, height, ventilation conditions, soil, crop, outside wind and temperature, solar radiation, etc. Then, these results are not directly extrapolated to others greenhouse conditions, but indicates that the heterogeneity conditions are able to occur in a normal way in typical greenhouses.

The differences in air temperature found for the different areas of our greenhouse appear excessive at certain times, possibly because of deficient central and lateral ventilation for this crop. Nevertheless, these excesses occur in normal cropping situations. Thus, the climatic conditions of a greenhouse are directly related to the outside climatic conditions and cultivation tasks, and under real operating conditions these pronounced temperature differences can occur regularly, getting the level considered as heterogeneous environment. Also, for other crops the heterogeneous conditions could be more accentuated. One clear example would be watermelon or melon cultivation in greenhouses during pollination tasks (Callejón-Ferre et al., 2009; 2011b). In this situation, the ventilation is closed to cause high humidity and high temperatures during the day, presumably with greater heterogeneity than in the present work.

Given our findings about the existence of heterogeneous conditions inside the greenhouse, further studies are required to evaluate the comfort and heat-stress indices. Although it is out of the scope of this work to demonstrate the existence of thermal stress, it is appropriate to adopt preventive measures as continued hydration during the working day and use of breathable light clothing (ILO, 2010; Jackson and Rosenberg, 2010; Stoecklin-Marois et al., 2013).

The UVI data show that, during the middle hours of the day inside the greenhouse, threshold risk values can be exceeded (UVI>2; Figure 8c). Regarding the maximum values, it seems clear that in the coming months of spring and summer the values would rise due not only with the higher position of the sun in the sky but also over time, due to the degradation of the plastic cover (López-Hernández, 2003). These facts show the importance of the sun protection, although to a lesser extent than in the outdoor agriculture (Kearney et al., 2013; Carley and Stratman, 2015). For this reason, it is recommended that workers use preventive measures such as sun cream, protective garments, and a wide-brimmed hat (WHO, 2002; Kearney et al., 2014) and manage their time to avoid the central hours of the day.

It bears mentioning that the number and placement of the measurement stations in the lanes of the greenhouse posed no problem for the routine cultivation tasks; the stations required only a protective cover of a plastic sheet during the biocide spraying.

Regarding the performance of the communication network, we collected an average of 3200 daily measurements of each sensor per station, which means a measurement every 20-30 seconds. A problem we observed is that, over the weeks, the radio signal quality (which is also logged into our dataset) progressively dropped as the plants grew. Clearly, the absorption of the 2.4 GHz signal by organic matter (in this case, the tomato plants) is at the origin of this problem. Eventually, some stations stopped giving reliable measurements via radio packets, so we used our mesh-like network design to redefine another logical tree to route the radio packets through different paths with less attenuation. In a couple of extreme cases, the radio emitter (XBee transceiver) had to be relocated, removing it from the electronic housing and placing it in the high part of the greenhouse.

**5. Conclusions**

In an Almería type greenhouse (S. Spain) a wireless network composed of 13 measurement stations was designed and manufactured and set at three heights throughout a greenhouse to collect climatic data related to the thermal environment in the workplace. Specifically, each measurement station was equipped with three sensors for air temperature, black globe temperature, relative humidity, and air velocity, situated at the three heights specified by ISO 7726 (i.e. ankle, abdomen, and head), plus an additional UVI radiation sensor.

The grid of twelve measurement stations distributed inside the greenhouse has demonstrated the existence of horizontal and vertical heterogeneity conditions with respect to air temperature according to the limits established by ISO 7726, specially on sunny days, where the temperature heterogeneity is especially relevant. The main outcome of our research is, therefore, that to be in accordance with guideline ISO 7726, thermal measurements aimed at labour studies in greenhouses should be taken at three heights and in different spatial placements along the greenhouse surface. Furthermore, UVI levels were found to be higher than WHO risk thresholds during the work day (UVI>2), thus making advisable to adopt preventive safety measures. Finally, we have validated our WSN design and proved that the mesh-like network topology represents an important advantage whenever plant growth reduces the radio transmissibility of some areas in the studied environment.

**Acknowledgements**

University Research Contract Number 401251 funded this study. Also to Laboratory-Observatory Andalusian Working Conditions in the Agricultural Sector (LASA).

**References**

Anastasi, G., Farruggia, O., Lo Re, G., Ortolani, M., 2009. Monitoring high-quality wine production using wireless sensor networks. In 42st Hawaii International International Conference on Systems Science (HICSS-42 2009): Waikoloa, Big Island, HI, USA.

Balendonck, J., Sapounas, A.A., Kempkes, F., Van Os, E.A., Van Der Schoor, R., Van Tuijl, B.A.J., Keizer, L.C.P., 2014. Using a wireless sensor network to determine climate heterogeneity of a greenhouse environment. Acta Hortic 1037, 539-546. doi: 10.17660/ActaHortic.2014.1037.67

Baronti, P., Pillai, P., Chook, V.W.C., Chessa, S., Gotta, A., Hu., Y.F., 2007. Wireless sensor networks: A survey on the state of the art and the 802.15.4 and ZigBee standards. Computer Communications 30, 1655-1695. doi:10.1016/j.comcom.2006.12.020

Budd, G.M., 2008. Wet-bulb globe temperature (WBGT) - its history and its limitations. J. Sci. Med. Sport 11, 20-32. doi:10.1016/j.jsams.2007.07.003

Cabrera-Sánchez, A., Uclés-Aguilera, D., Agüera-Camacho, T., De la Cruz-Fernández, E., 2016. Informes y Monografías/54. Análisis de la campaña hortofrutícola de Almería 2015/2016. [Reports and monographs/54. Analysis of the fruit and vegetable sector of Almería 2015/2016], Ed. Cajamar Caja Rural, Almería, Spain. <http://www.publicacionescajamar.es/pdf/series-tematicas/informes-coyuntura-analisis-de-campana/analisis-de-la-campana-hortofruticola-17.pdf>, (accessed 28.06.17).

Callejon-Ferre, A.J., Manzano-Agugliaro, F., Diaz-Perez, M., Carreno-Sanchez, J., 2011a. Improving the climate safety of workers in Almería-type greenhouses in Spain by predicting the periods when they are most likely to suffer thermal stress. Appl. Ergon. 42, 391-396. doi:10.1016/j.apergo.2010.08.014

Callejón-Ferre, A.J., Pérez-Alonso, J., Carreño-Ortega, A., Velázquez-Martí, B., 2011b. Indices of ergonomic-psychosociological workplace quality in the greenhouses of Almería (Spain): crops of cucumbers, peppers, aubergines and melons. Safety Sci. 49, 746-750. doi:10.1016/j.ssci.2010.12.009

Callejón-Ferre, A.J., Pérez-Alonso, J., Sánchez-Hermosilla, J., Carreño-Ortega, A., 2009. Ergonomics and psycho-sociological quality indices in greenhouses Almería (Spain). Span. J. Agric. Res. 7, 50-58.

Carley, A., Stratman, E., 2015. Skin Cancer Beliefs, Knowledge, and Prevention Practices: A Comparison of Farmers and Nonfarmers in a Midwestern Population. J. Agromedicine 20, 85-94. doi:10.1080/1059924X.2015.1010059

Carmona-Benjumea, A., 2001. Datos antropométricos de la población laboral española. Informe de resultados. Seguridad y Salud en el Trabajo 14, 22-35. <http://www.insht.es/portal/site/Insht/menuitem.1f1a3bc79ab34c578c2e8884060961ca/?vgnextoid=b56aaddc77d77110VgnVCM100000b80ca8c0RCRD&vgnextchannel=9f164a7f8a651110VgnVCM100000dc0ca8c0RCRD>, (accessed 28.06.2017). [in Spanish].

Castilla, N., 2005. Invernaderos de plástico. Tecnología y manejo, Mundi-prensa, Madrid. [in Spanish].

Cecchini, M., Colantoni, A., Massantini, R., Monarca, D., 2010. Estimation of the risks of thermal stress due to the microclimate for manual fruit and vegetable harvesters in central Italy. J. Agric. Safety Health 16, 141-159.

D'Ambrosio-Alfano, F.R., Malchaire, J., Palella, B.I., Riccio, G., 2004. WBGT index revisited after 60 years of use. Ann. Occup. Hyg. 58, 955-970 doi:10.1093/annhyg/meu050

D'Ambrosio-Alfano, F.R., Palella, B.I., Riccio, G., 2011. Thermal environment assessment reliability using temperature -Humidity indices. Ind. Health 49, 95-106.

Diano, M., Valentini, M., Samele, P., Di Gesu, I., 2016. Exposure to hot environments of horticultural greenhouses workers of the center of Calabria: evaluative comparison methods. Italian J. Occupational Environmental Hygiene 7, 56-114.

Dursch, A., Yen, D.C., Shih, D.H., 2004. Bluetooth technology: an exploratory study of the analysis and implementation frameworks. Computer Standards and Interfaces 26, 263-277. doi:10.1016/j.csi.2003.12.005

Epstein, Y., Moran, D.S., 2006. Thermal Comfort and the Heat Stress Indices. Ind. Health 44, 388-398.

Ferentinos, K.P., Katsoulas, N., Tzounis, A., Bartzanas, T., Kittas, C., 2017. Wireless sensor networks for greenhouse climate and plant condition assessment. Biosys. Engin 153, 70-81. doi: 10.1016/j.biosystemseng.2016.11.005

Fernández, C., Pérez, J.J., 2004. Caracterización de los invernaderos de la provincia de Almería. Ed. Cajamar. Almería, Spain. [in Spanish].

Granados, M.R., López, J.C., Bonachela, S., Hernández, J., Magán, J.J., 2016. Perfiles de temperatura en invernadero con acolchado negro y cultivo de pepino en periodos fríos. II Simposio Nacional de Ingeniería Hortícola. Almería, Spain. [in Spanish].

Hamrita, T.K., Hoffacker, E.C., 2005. Development of a "smart" wireless soil monitoring sensor prototype using RFID technology. Appl. Eng. Agric. 21, 139-143.

IEEE. 2009. IEEE802.15: "Part 15.4: Wireless Medium Access Control (MAC) and Physical Layer (PHY) Specifications for Low-Rate Wireless Personal Area Networks

(WPANs)", IEEE Std 802.15.4d-2009. <http://standards.ieee.org/getieee802/download/802.15.4d-2009.pdf>, (accessed 28.06.2017)

IEEE. 2012. IEEE 802.11: Wireless LAN Medium Access Control (MAC) and Physical Layer (PHY) Specifications. IEEE-SA. doi:10.1109/IEEESTD.2012.6178212

ILO (International Labour Office), 1985. R171-Occupational Health Services Recommendation. Adoption: Geneva, 71st ILC session (26 Jun 1985). <http://www.ilo.org/dyn/normlex/en/f?p=1000:12100:0::NO::P12100_INSTRUMENT_ID,P12100_LANG_CODE:312509,es:NO>, (accessed 28.06.17).

ILO (International Labour Office), 2010. Code of practice on safety and health in agriculture. Sectoral Activities Programme. Meeting of Experts to Adopt a Code of Practice on Safety and Health in Agriculture, Geneva, Switzerland. <http://www.ilo.org/wcmsp5/groups/public/---dgreports/---dcomm/---publ/documents/publication/wcms_159457.pdf>, (accessed 28.06.17).

ISO 17166, 1999. Erythema reference action spectrum and standard erythema dose. International Organization for Standardization, Geneva, Switzerland.

ISO 7243, 1989. Hot environments - estimation of the heat stress on working man based on the WBGT-index (Wet Bulb Globe Temperature), International Organization for Standardization, Geneva, Switzerland.

ISO 7726, 1998. Ergonomics of the thermal environment and instruments for measuring physical quantities, International Organization for Standardization, Geneva, Switzerland.

ISO 7730, 2005. Ergonomics of the thermal environment – analytical determination and interpretation of thermal comfort using calculation of the PMV and PPD indices and local thermal comfort, International Standardization Organization, Geneva, Switzerland.

ISO 7933, 2004. Ergonomics of the thermal environment: Analytical determination and interpretation of heat stress using calculation of the predicted heat strain, International Organization for Standardization, Geneva, Switzerland.

ISO 8996, 2004. Ergonomics of the thermal environment -- Determination of metabolic rate, International Organization for Standardization, Geneva, Switzerland.

ISO 9920, 2007. Ergonomics of the thermal environment -- Estimation of thermal insulation and water vapour resistance of a clothing ensemble, International Organization for Standardization, Geneva, Switzerland.

Jackson, L.L., Rosenberg, H.R., 2010. Preventing Heat-Related Illness Among Agricultural Workers. J. Agromedicine, 15, 200-215. doi:10.1080/1059924X.2010.487021

Kearney, G.D., Gregory D., Lea, C.S., Balanay, J., Wu, Q., Bethel, J.W., Von Hollen, H., Sheppard, K., Tutor-Marcom, R., Defazio, J., 2013. Assessment of Sun Safety Behavior among Farmers Attending a Regional Farm Show in North Carolina. J. Agromedicine 18, 65-73. doi: 10.1080/1059924X.2012.743378

Kearney, G.D., Kearney, Gregory D., Xu, X.H., Balanay, J.A.G., Becker, A.J., 2014. Sun Safety Among Farmers and Farmworkers: A Review. J. Agromedicine 19, 53-65. doi:10.1080/1059924X.2013.855691

López, A. Valera, D.L., Molina-Aiz, F.D., Peña, A., 2012a. Sonic anemometry to evaluate airflow characteristics and temperature distribution in empty Mediterranean greenhouses equipped with pad-fan and fog systems. Biosys. Engin. 113, 334-350. doi: 10.13031/2013.30077

López, A., Valera, D.L., Molina-Aiz, F.D., Peña, A., 2012b. Sonic anemometry measurements to determine airflow patterns in multi-tunnel greenhouse. Span. J. Agric. Res. 10, 631-642. doi: 10.5424/sjar/2012103-660-11

López, A., Valera, D.L., Molina-Aiz, F.D., Peña, A., 2013. Effectiveness of horizontal air flow fans supporting natural ventilation in a Mediterranean multi-span greenhouse. Sci. Agric. 70, 219-228. doi: 10.1590/S0103-90162013000400001

López-Hernánez, J.C., 2003. Evolución de estructuras y cubiertas de invernadero del sureste de español. Técnicas de producción en cultivos protegidos. Tomo I. Ed. Cajamar. Almería, Spain. [in Spanish].

López Riquelme, J.A., Soto, F., Suardíaz, J., Sánchez, P., Iborra, A., Vera, J.A., 2009. Wireless Sensor Networks for precision horticulture in Southern Spain. Comput Electron Agr 68, 25-35. doi: 10.1016/j.compag.2009.04.006

Marucci, A., Pagniello, B., Monarca, D., Cecchini, M., Colantoni, A., Biondi, P., 2012. Heat stress suffered by workers employed in vegetable grafting in greenhouses. J. Food Agric. Environ. 10, 1117-1121.

Masterton, J., Richardson, F.A., 1979. Humidex, a Method of Quantifying Human Discomfort Due to Excessive Heat and Humidity, Environment Canada, Downsview, Ontario, Canada.

Molina-Aiz, F.D., Valera, D.L., Álvarez, A.J., 2004. Measurement and simulation of climate inside Almería-type greenhouses using computational fluid dynamics. Agric. For Meteorol. 125, 33-51. doi: 10.1016/j.agrformet.2004.03.009

Nijskens, J., Deltour, J., Coutisse, S., Nisen, A., 1985. Radiation transfer through covering materials, solar and thermal screens of greenhouses. Agr. Forest Meteorol. 35, 229-242. doi:10.1016/0168-1923(85)90086-3

Okushima, L., Sase, S., Lee, I.B., Bailey, B.J., 2001. Thermal environment and stress of workers in naturally ventilated greenhouses under mild climate. Acta Hortic. 559, 793-798. doi:10.17660/ActaHortic.2001.559.118

Parsons, K., 2013. Occupational health impacts of climate change: Current and future ISO standards for the assessment of heat stress. Ind. Health 51, 86-100.

Pérez-Alonso, J., Callejón-Ferre, A.J., Carreño-Ortega, A., Sánchez-Hermosilla, J., 2011. Approach to the evaluation of the thermal work environment in the greenhouse-construction industry of SE Spain. Build. Environ. 46, 1725-1734. doi: 10.1016/j.buildenv.2011.02.014

Rohles, F.H., 1985. Environmental ergonomics in agricultural systems. Appl. Ergon. 16, 163-166. doi: 10.1016/0003-6870(85)900002-X

Ruiz-García, L., Lunadei, L., Barreiro, P., Robla, J.I., 2009. A review of wireless sensor technologies and applications in agriculture and food industry: state of the art and current trends. Sensors 9, 4728-4750. doi:10.3390/s90604728

Stoecklin-Marois, M., Hennessy-Burt, T., Mitchell, D., Schenker, M., 2013. Heat-related Illness Knowledge and Practices among California Hired Farm Workers in The MICASA Study. Ind. Health, 51, 47-55. doi: http://doi.org/10.2486/indhealth.2012-0128

Srbinovska, M., Gavrovski, C., Dimcev, V., Krkoleva, A., Borozan, V., 2015. Environmental parameters monitoring in precision agriculture using wireless sensor networks. J Clean Prod 88, 297-307. doi: 10.1016/j.jclepro.2014.04.036

Vox, G., Losito, P., Valente, F., Consoletti, R., Scarascia-Mugnozza, G., Schettini, E., Marzocca, C., Corsi, F., 2014. A wireless telecommunications network for real-time monitoring of greenhouse microclimate. Journal of Agricultural Engineering 45, 70-79. doi: 10.4081/jae.2014.237

WHO (Worl Health Organization), 2002. Global solar UV index: a practical guide. A joint recommendation of the World Health Organization, World Meteorological Organization, United Nations Environment Programme and the International Commission on Non-Ionizing Radiation Protection. World Health Organization, Geneva, Switzerland. <http://www.who.int/uv/publications/en/GlobalUVI.pdf?ua=1>, (accessed 28.06.17).